\documentclass[aps,prl,twocolumn,nopacs,superscriptaddress,longbibliography]{revtex4-1}

\usepackage[breaklinks=true,colorlinks,citecolor=blue,linkcolor=blue,urlcolor=blue]{hyperref}
\usepackage{epsfig,mathrsfs,xcolor,latexsym,subfigure,marginnote,graphicx,verbatim,relsize,mathrsfs,color,array,amsmath,amsfonts,amssymb,graphicx}
\usepackage{braket}

\begin{document}

\title{Peierls distortion driven multi-orbital origin of charge density waves in the undoped infinite-layer nickelate}

\author{Ruiqi~Zhang}
\affiliation{Department of Physics and Engineering Physics, Tulane University, New Orleans, LA 70118, USA}

\author{Christopher Lane}
\email{laneca@lanl.gov}
\affiliation{Theoretical Division, Los Alamos National Laboratory, Los Alamos, New Mexico 87545, USA}

\author{Johannes Nokelainen}
\affiliation{Department of Physics, Northeastern University, Boston, MA 02115, USA}

\author{Bahadur Singh}
\affiliation{Department of Condensed Matter Physics and Materials Science, Tata Institute of Fundamental Research, Colaba, Mumbai 400005, India}

\author{Bernardo~Barbiellini}
\affiliation{Department of Physics, School of Engineering Science, LUT University, FI-53850 Lappeenranta, Finland}
\affiliation{Department of Physics, Northeastern University, Boston, MA 02115, USA}

\author{Robert S. Markiewicz}
\affiliation{Department of Physics, Northeastern University, Boston, MA 02115, USA}

\author{Arun~Bansil}
\email{ar.bansil@neu.edu}
\affiliation{Department of Physics, Northeastern University, Boston, MA 02115, USA}

\author{Jianwei~Sun}
\email{jsun@tulane.edu}
\affiliation{Department of Physics and Engineering Physics, Tulane University, New Orleans, LA 70118, USA}

\begin{abstract}

Understanding similarities and differences between the cuprate and nickelate superconductors is drawing intense current interest. Competing charge orders have been observed recently in the $undoped$ infinite-layer nickelates in sharp contrast to the $undoped$ cuprates which exhibit robust antiferromagnetic insulating ground states. The microscopic mechanisms driving these differences remain unclear. Here, using in-depth first-principles and many-body theory based modeling, we show that the parent compound of the nickelate family, LaNiO$_2$, hosts a charge density wave (CDW) ground state with the predicted wavevectors in accord with the corresponding experimental findings. The CDW ground state is shown to be connected to a multi-orbital Peierls distortion. Our study points to the key role of electron-phonon coupling effects in the infinite-layer nickelates.

\end{abstract} 

\maketitle

\section{Introduction}

Recent discovery of superconductivity in the infinite-layer nickelates~\cite{Li2019a,Osada2020a,ZhangS2021,2021arXiv211202484R,OsadaAM2021} has attracted intense attention~\cite{Sawatzky2019,Norman2020,Pickett2021,Mitchell2021} as an analog of cuprate superconductivity. Much experimental and theoretical effort has been devoted to understanding the relationship between the nickelates and cuprates ~\cite{Hepting2020,Lee2004,Zhang2020d,Sakakibara2019,Botana2020,Goodge2021,Jiang2020,Zhang2021}. Experiments have found strong hints of magnetism in the {\it undoped} infinite-layer nickelates ~\cite{Lu2021science,Fowlie2022,Lin_2022,OrtizPRR2022}, similar to the case of the cuprates. Strikingly, evidence of charge density wave (CDW) order has also been reported in {\it undoped} nickelates~\cite{2021arXiv211202484R,Krieger2021,Tam2021}, which differs drastically from the {\it undoped} cuprates which exhibit a well-defined AFM ground state.  

Although the CDWs in the infinite-layer nickelates were reported by several groups almost simultaneously~\cite{2021arXiv211202484R,Krieger2021,Tam2021}, the mechanism driving the CDW order remains unclear. Ref.~\cite{2021arXiv211202484R} suggests that the CDWs could be driven by strong correlation effects in the infinite-layer nickelates. But, ref.~\cite{Tam2021} suggests that charge modulations arise from electron-phonon (\textit{e-ph}) coupling. Ref.~\cite{Gao2022magstrain} also points to the significance of \textit{e-ph} coupling effects in the nickelates. Krieger \textit{et. al.}~\cite{Krieger2021}, however, suggest that the CDW in the nickelates may be related to the capping layer, although robust CDW signals have been reported in non-capped nickelates in ref.~\cite{Tam2021}. These contradictory results must be reconciled and the mechanism driving the appearance of the CDW order in the nickelates must be identified before we can get a handle on the rich phase diagram of the nickelates.

Here, we present a comprehensive theory of charge orders in the {\it undoped} infinite-layer nickelates. Our first principles density functional theory (DFT) computations demonstrate that, much like the $doped$ cuprates but in sharp contrast to the $undoped$ cupates~\cite{Zhang2020}, the $undoped$ infinite-layer nickelate hosts a rich landscape of competing low-energy stripe and magnetic phases. Moreover, our many-body theory based modeling finds the existence of incommensurate stripe phases that are not accessible to the DFT, indicating the presence of strong electron correlation effects in the parent compound of the nickelates, in line with recent experiments~\cite{Lu2021science,Fowlie2022,Lin_2022,OrtizPRR2022}.

\begin{figure*}[htpb]
	\centering
	\includegraphics[width=0.8\linewidth]{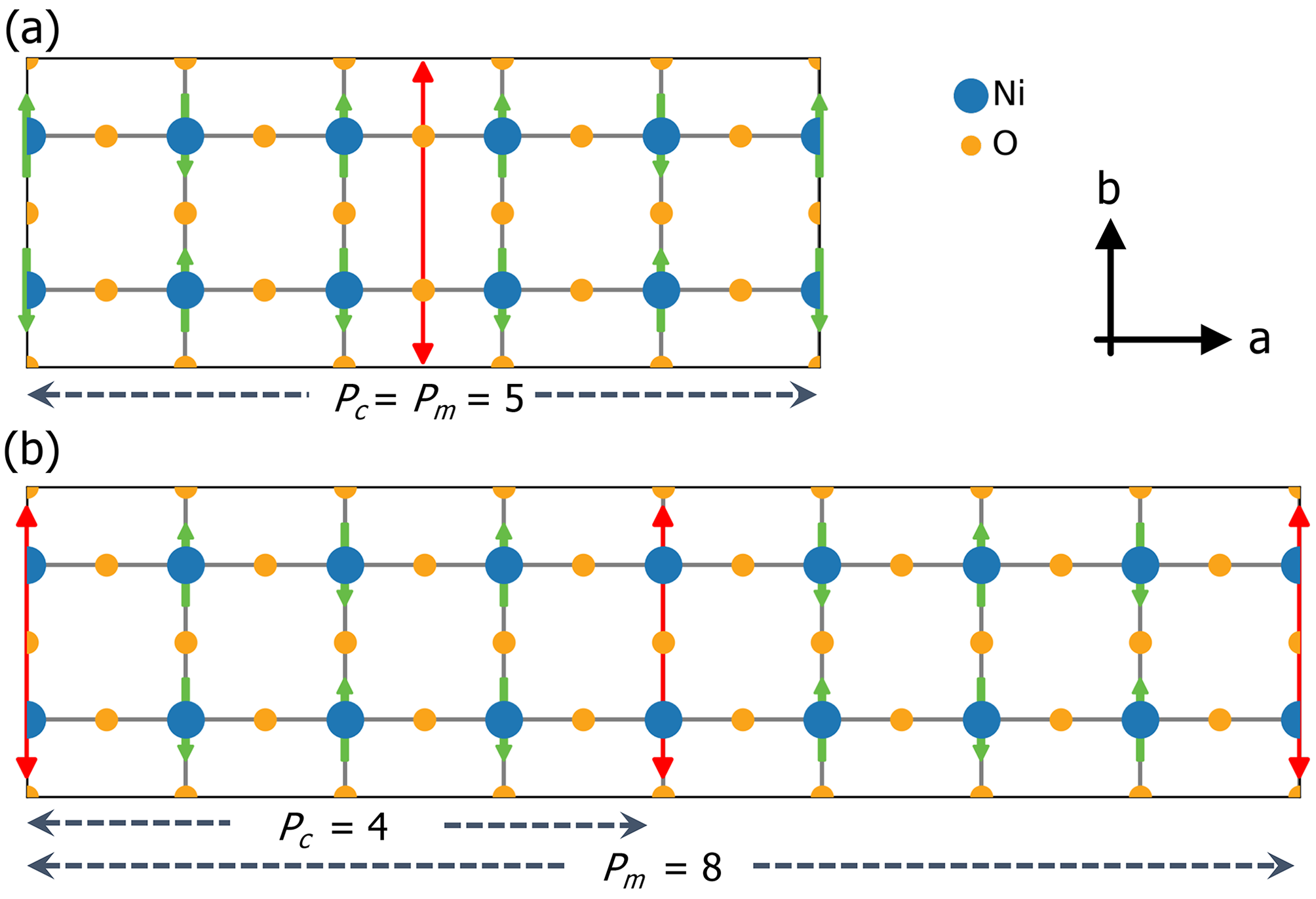}
	\caption{{\bf Schematics of two 1D stripe phases along the $a$-axis in LaNiO$_{2}$.} (a) A bond-centered stripe phase, where the antiphase boundaries (APBs) pass through the Ni--O--Ni bond, i.e., are centered on the bonding oxygen atoms. Here $P_{c} = P_{m} = 5$, where $P_{c}$ and $P_{m}$ denote the periodicities of the charge and spin orders, respectively.  (b) A site-centered stripe phase with $P_{m} =8$ and $P_{c} = 4$, where the APB passes through the Ni atoms. The blue and yellow circles represent Ni and O atoms, respectively. La atoms are omitted in the figure for clarity. The green arrows indicate the relative orientation of magnetic moments on Ni atoms, and the red lines with  double arrows point to the APBs.} 
	\label{fig:fig1} 
\end{figure*}

Strikingly, the predicted CDW vectors of the stripe phase with the lowest energy in our calculations are in excellent accord with the recent experimental observations~\cite{2021arXiv211202484R,Krieger2021,Tam2021}. We find that a number of Ni $d$ orbitals conspire to stabilize the CDW in the nickelates. This stabilization is connected to Peierls distortion of mainly the Ni $d$ orbitals, where the stripes open minigaps in the Ni $3d_{z^2}$ and $3d_{xz/yz}$ bands. This finding shows that (\textit{e-ph}) coupling is crucial for understanding the nickelates, as suggested by a recent experiment~\cite{Gao2022magstrain}. In addition, the large gap in the Ni $3d_{x^2-y^2}$ band makes a significant contribution to the spin moment, but leaves the $3d_{x^2-y^2}$ electrons far from the Fermi level, similar to their role in {\it undoped} cuprates. Furthermore, in both the $C$-AFM and stripe phases of the nickelates, the calculated band structures contain flat bands dominated by Ni $3d_{z^2}$ orbitals near the Fermi level, inducing Van Hove singularities in the density of states (DOS), which could enhance many-body and \textit{e-ph} interactions and drive more exotic correlated phenomena~\cite{Zhang2021,ChoiPRR2020}. Our findings, therefore, provide insight into the low-energy physics, the lack of long-range magnetic order, and the origin of the CDW in nickelates. These results indicate that not only strong correlation effects but also \textit{e-ph} coupling is of importance for nickelates, and a one band model based on $d_{x^2-y^2}$ is not sufficient to describe low-energy physics of LaNiO$_{2}$.

\section{Results}
\subsection{Stripe Phases in LaNiO$_2$}

We have systematically performed calculations on a number of 1D stripe phases as a function of charge periodicity. We begin by considering two exemplar stripe phases that are schematized in Fig.~\ref{fig:fig1}. We assume that the stripe phases have the same ferromagnetic inter-layer exchange couplings as in the $C$-AFM phase. Firstly, we study a bond-centered stripe phase with charge periodicity $P_{c} = 5$ and magnetic periodicity $P_{m} = 5$, see Fig.~\ref{fig:fig1} (a). The calculated magnetic moments of Ni atoms are around 1.06\,$\mu_{B}$ with small modulations of $\sim$0.02$\,\mu_{B}$, leading to a slightly enhanced magnetic density around the antiphase boundaries (APBs). This phenomenon is discussed in further detail in the following section. Secondly, we study a site-centered stripe phase with charge and magnetic periodicities of $P_{c} = 4$ and $P_{m}= 8$, see Fig.~\ref{fig:fig1}\,(b). Notably, we find the self-consistent magnetic moment of each Ni atom along the APBs to be 0.00\,$\mu_{B}$, whereas the Ni atoms away from the APB exhibit a 0.99\,$\mu_{B}$ moment in site-centered cases. This difference between bond- and site-centered cases persists for stripe phases within the more comprehensive set of stripe phases that we have studied. This set contains both bond- and site-centered cases for $P_{c} = 3$, 4, 5, 6, 7, 9, 11, 13, and 15. Computations for larger values of $P_{c}$ become impractical. Next, we discuss these results in more detail. 

\begin{figure}[t]
	\centering
	\includegraphics[width=0.99\columnwidth]{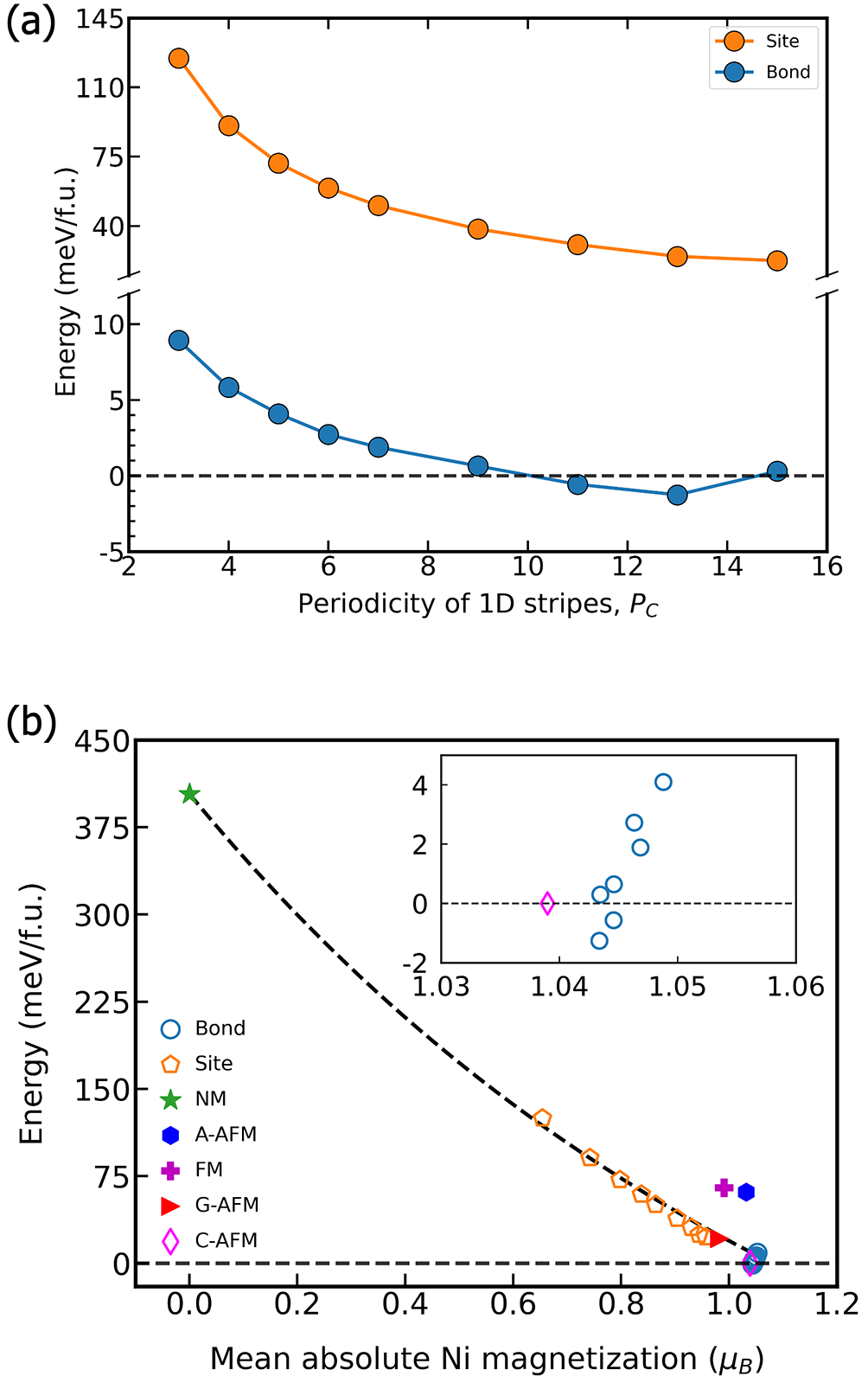}
	\caption{{\bf Stability of the various computed states in LaNiO$_{2}$.}  (a) Relative energies of a number of stripe phases with respect to the $C$-AFM state are plotted as a function of the periodicity of the various 1D charge stripes.  (b) Energies of the computed states of LaNiO$_{2}$ relative to the $C$-AFM structure as a function of the average magnetic moment of Ni atoms are shown. The dashed black line is a guide to the eye. The insert is a blow-up of the lower right corner of the plot.}
	\label{fig:fig2}
\end{figure}

Figure~\ref{fig:fig2} (a) shows the total energies of each site and the bond-centered stripe state for undoped LaNiO$_{2}$ relative to the $C$-AFM phase~\cite{Zhang2021}. Our total energy calculations show the bond-centered stripe is more stable than the site-centered stripe, consistent with results on cuprates~\cite{Zhang2020}. Moreover, we find that the bond-centered stripes become more stable as the charge periodicity increases, consistent with pristine YBa$_2$Cu$_3$O$_{6}$ (YBCO$_6$)~\cite{Zhang2020}. Figure~\ref{fig:fig2} (b) compares the total energy of the various magnetic phases with the average Ni magnetization. Here, we include all of the 1D stripe phases up to $P_{c} = 15$ and the $C$-AFM phase as the reference, along with a variety of uniform magnetic phases that have been previously discussed in Ref. ~\cite{Zhang2021}. Overall, we find that a larger average Ni magnetic moment results in a more stable magnetic phase and that several high-magnetic-moment phases are nearly degenerate in energy. Two of the bond-centered stripe phases ($P_{c} = 11$ and $P_{c} = 13$) are found to be more stable than the $C$-AFM phase. In particular, multiple phases with strong local magnetic moments are found to be nearly degenerate in energy, with quite small energy differences of less than 2 meV. The bond-centered stripe phase with $P_{c} = 13$ is found to be the ground state. These results are in remarkable agreement with our recent study of the yttrium-based cuprates~\cite{Zhang2020}, which revealed that ``intertwined orders" with larger magnetic moments are crucial for understanding the properties of correlated materials. Notably, we rule out any significant role of the non-magnetic (NM) phase in the ground state since it is not energetically competitive ($\sim$400 meV above the $C$-AFM phase). We emphasize that, similar to the doped cuprates, Fig.~\ref{fig:fig2} clearly provides evidence for the existence of competing low-energy states in LaNiO$_2$.

The computational cost becomes prohibitive for larger $P_c$ values using DFT calculations. To explore the full manifold of possible low-energy stripe phases, we perform a random phase approximation (RPA) based instability analysis of the NM state [Section 1 of  Supplementary Information(SI)]. The RPA results are consistent with DFT, showing that the total energy of the nickelate system is lowered by hosting a number of nearly degenerate stripe phases.

\begin{figure*}[htpb]
	\centering
	\includegraphics[width=0.98\linewidth]{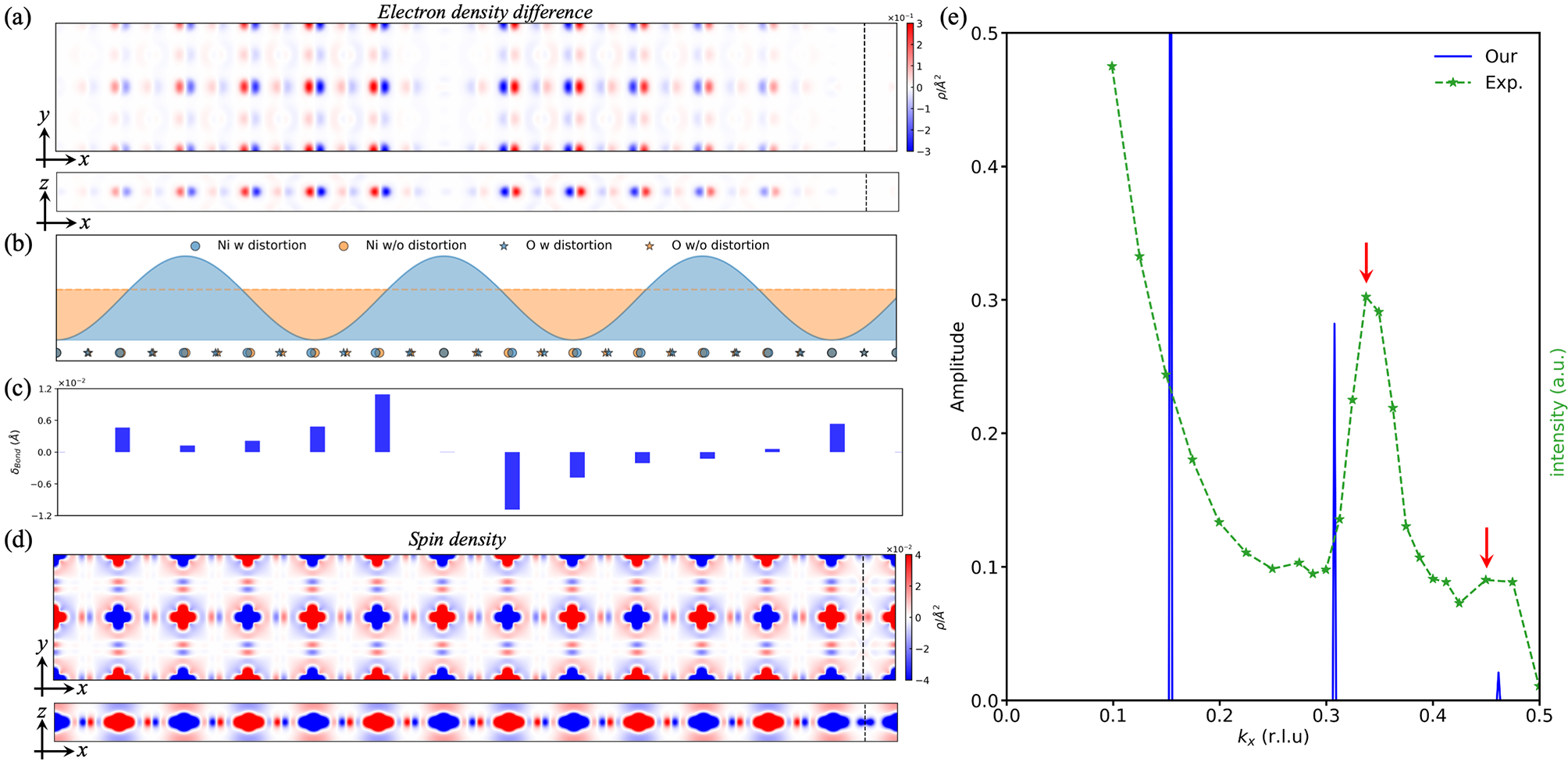}
	\caption{{\bf Charge and spin redistribution, lattice distortions, and 1D Fourier transformation analysis.} 
 The top and side view of the projection of (a) electron density difference between the stripe phase with $P_m = P_c = 13$ and the $C$-AFM phase on the $xy$-plane. Red indicates electron accumulation and blue depletion. (b) (top) Schematic plot of the $k_x=4\pi/13a$ CDW in the stripe phase with $P_m = P_c = 13$. (bottom) The blue and yellow balls represent the positions of Ni atoms with and without distortion, respectively, while the blue and yellow stars plot the corresponding positions of O atoms with and without distortion. To show the displacement of atoms clearly, we enlarge the original values from DFT calculations by ten times.  (c) The bond length difference bar plot for the right and left side Ni-O bonds of each Ni atom. (d) The top and side view of spin density of bond-centered stripe phase with $P_m = P_c = 13$. Here red and blue denote spin-up and spin-down, respectively. The antiphase boundary is denoted by a black dashed line. (e) The 1D Fourier transformation of charge density (a) of the stripe phase with $P_m = P_c = 13$ along $x$ direction. The amplitude values were normalized in the first B.Z. The intensity of the RIXS quasi-elastic data of undoped LaNiO$_2$ is adapted from ref.~\cite{2021arXiv211202484R} to  compare with our prediction.}
	\label{fig:fig3}
\end{figure*}

Based on the DFT total energy calculations, we find one significant difference between the cuprates and nickelates: in the undoped cuprates, the stripe phases are always metastable, with excess energy $\sim$$1/P_{c}$, leading to a well-defined AFM ground state~\cite{Zhang2020}, whereas in the nickelates a stripe phase is the lowest energy state similar to YBCO$_7$~\cite{Zhang2020}. 

This suggests that the nickelates may be closer to the Slater regime, where Fermi surface nesting can lower the stripe energy.  This enhanced stripe-AFM competition may explain why LaNiO$_{2}$, like YBCO$_{7}$, has no long-range magnetic order~\cite{OrtizPRR2022,Li2019a}. Indeed, the similarity of Fig.~2(b) to Fig.~3 of ref.~\cite{Zhang2020}, with both displaying an accumulation point of phases near the ground state, is striking, and it provides a plausible basis for the intertwined orders found in many correlated materials.

\subsection{Charge Redistribution}
\label{C_R}

To understand the stabilization of these stripe phases, we plot selected charge differences between the bond-centered stripes and $C$-AFM phases
in Figs.~\ref{fig:fig3} (a) and S2 in section 2 of SI. For $P_c \leq 9$ phases, strong charge redistribution is found around the APBs, which minimizes the effect of parallel neighboring magnetic configurations on the boundaries. However, for $P_c \geq 11$, we find that the charges are deconfined from the APBs, predominantly redistributing in the magnetic domains, which leads to $P_c = 11$ and $13$ showing lower energies than the $C$-AFM phase. Taking $P_c = 13$ as an example, in  Fig.~\ref{fig:fig3} (a) we plot the electron density difference -- the $P_c=13$ charge density minus the $C$-AFM charge density stretched to have the same lattice constant.  We find a strong charge polarization along the $x$ direction, but not along the $z$ direction, where the electron clouds around each Ni or O atom are polarized to point away from a depletion region centered half way between the two neighboring antiphase boundaries.  Consistent with this, all of the Ni and O atoms are displaced away from the depletion layer, Fig.~\ref{fig:fig3} (b), bottom, the accompanying Peierls distortion, leading to a more stable stripe phase than $C$-AFM. [Note, displacements are enhanced by a factor of 10 for ease of viewing.]  Due to unequal atomic displacements, the Ni atoms are shifted away from the midpoint between the two neighboring O atoms, Fig.~\ref{fig:fig3} (c), leading to additional polarization effects. To relate this to the observed CDW, in the top part of Fig.~\ref{fig:fig3} (b) we sketch a CDW of the correct periodicity.  We note that the shape of the central lobe closely matches the profile of the depletion layer, as seen from either the electron polarization (a) or the atomic displacements (b), while the side lobes are not in evidence in the data.

In Fig.~\ref{fig:fig3} (d), we show the spin density of the bond-centered stripe phase with $P_c = 13$. From the shape of the Ni orbitals, one can clearly see the local magnetic moments of Ni sites are mainly composed of Ni $3d_{x^2-y^2}$ and $3d_{z^2}$ orbitals. We also note that O atoms in the magnetic domains show small spin polarization, while O atoms around the APBs show a dumbbell spin density, which implies that they have a small magnetic moment, 0.037$\mu_B$. These findings strongly indicate that nickelates are not a charge-transfer system like the cuprates~\cite{Zhang2020}. 

Figure~\ref{fig:fig3} (e) shows the 1D Fourier transform (FT) of the charge density of the bond-centered stripe phase with $P_c = P_m = 13$ along the $x$ direction. We identify three peaks in the first Brillouin Zone (B.Z.). Note that although the first peak located at 0.15 has the strongest amplitude, its wavelength may overlap with the tail of the Q=0 peak in experiments. Importantly, the wave vector of our second peak is 4/13 = 0.31, which is very close to the  experimental values of 0.33~\cite{2021arXiv211202484R,Krieger2021,Tam2021}. In addition, a weaker peak around 6/13 = 0.47 is noted in both our calculations and experiment~\cite{2021arXiv211202484R}. While the charge transfer between the sites is  small [Fig. S3], there is a significant charge polarization associated with the Peierls distortion [Figs.~\ref{fig:fig3} (a) and (b)], which presumably drives CDW formation.

\subsection{Electronic Structures}

\begin{figure*}[htpb]
	\centering
	\includegraphics[width=0.95\linewidth]{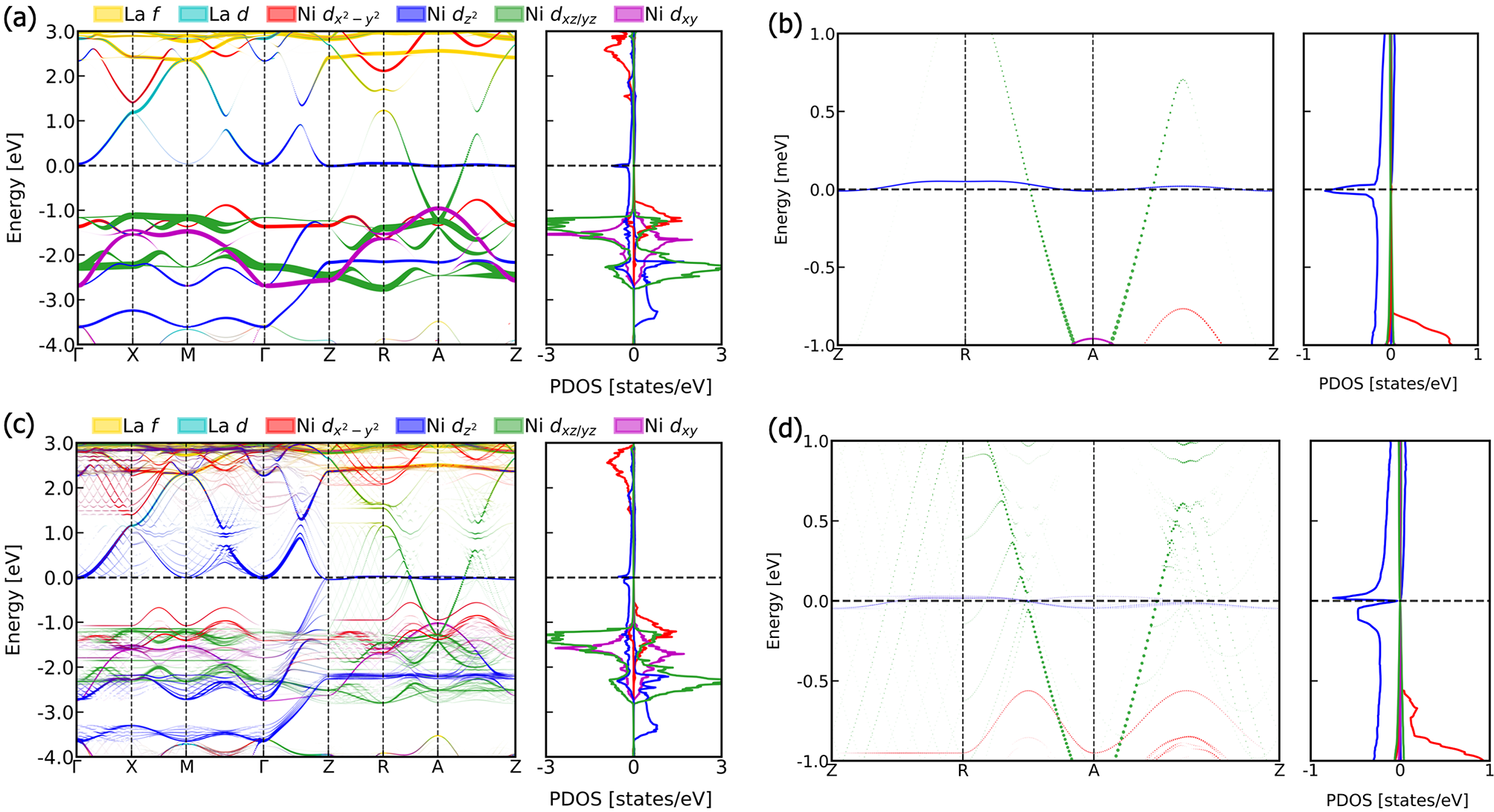}
	\caption{{\bf Comparison of the electronic structures of $C$-AFM and bond-centered stripe phases.} (a) Orbitally-projected electronic structures of $C$-AFM phase. The site-projected orbitally-resolved partial densities of states are for Ni sites with positive magnetic moments. Positive (negative) values mean spin-up (spin-down) DOS.  (b) is the blow-up of (a) with an energy window from -1 to 1 eV. (c) Same as (a), but for the bond-centered phase with $P_{c}=P_{m}= 13$. (d) is the blow-up of (c) with an energy window from -1 to 1 eV.  Band structures in (a) and (c) are ``unfolded” onto the primitive $1\times 1 \times 1$ Brillouin zone, with the intensity proportional to the spectral weight.}
	\label{fig:fig4}
\end{figure*}

To gain insight into the electronic properties of the ground state, we compare the electronic band structures of the ground state $P_{c} =P_{m} = 13$ bond-centered stripe and the uniform $C$-AFM phase. Figure~\ref{fig:fig4} presents the unfolded electronic band structures of LaNiO$_2$ in the $C$-AFM (a) and $P_{c} =P_{m} = 13$ bond-centered stripe (c) phases, respectively. As pointed out in Ref.~\cite{Zhang2021}, the magnetic state appears as a Hund's phase, where the magnetic moments primarily consist of $d_{x^2-y^2}$ character (0.75$\mu_B$), with a smaller admixture from $d_{z^2}$ (0.25$\mu_B$). This is reflected in the band structure. The Ni-$d_{x^2-y^2}$ band (red) is fully gapped out, consistent with recent experiment~\cite{Tam2021}, and the $d_{z^2}$ is spin-split (blue), similar to the cuprates~\cite{ChrisLCO2018}. Due to the fractional filling of the upper Ni-$d_{z^2}$ band, it is pinned to the Fermi level producing a flat band in the $k_z=\pi$ plane, consistent with previous studies~\cite{Zhang2021,ChoiPRR2020}. Flat bands dominated by Ni-$d_{z^2}$ have also been predicted in multilayer nickelates using DFT+DMFT calculations~\cite{LechermannPRB2022}. Similar results are found for the $P_c=P_{m}=13$ stripe (0.78 $\mu_B$ from $d_{x^2-y^2}$, 0.26$\mu_B$ from $d_{z^2}$). Remarkably, this pinning is absent in the NM phase, and arises from a strong reorganization of the $d_{z^2}$ dispersion when a magnetic gap opens on the $d_{x^2-y^2}$ band. This flat band gives rise to a Van Hove singularity (VHS) in the DOS, which can enhance the many-body interaction and drive more exotic correlated phenomena~\cite{Robert2021}. This flat band has also been suggested to couple electronic and lattice instabilities~\cite{Zhang2021,ChoiPRR2020}. Note that the VHS in LaNiO$_2$, involving the Ni $3d_{z^2}$ orbital, is stronger than the $3d_{x^2-y^2}$ VHS found in cuprates. However, it is extremely rare to find a VHS exactly at the Fermi level, without driving a transition to a new phase.

Upon unfolding the unit cell to capture the $P_{c} = P_{m} = 13$ stripe, we find the resulting band dispersions [Figs.~\ref{fig:fig4}(c)] are only slightly modified from those of the $C$-AFM phase. Specifically, the degeneracy of bands in the stripe phase gets lowered due to the reduction in symmetry. Despite the lower symmetry, the 3$d_{z^2}$ flat band remains pinned to the Fermi level, as is also found in other stripe phase, as we show in SI, section 4. Away from the Fermi level, minigaps form in the Ni-$3d_{z^2}$/La-$d$ hybridized band due to the supermodulation of the crystal potential, typical of stripe-like structures~\cite{MarkiewiczPRB2000}. This is also consistent with cuprates, where the stripes were identified as topological defects of an underlying AFM order. From Fig.~\ref{fig:fig4}(d), we find clear evidence that the stripe stabilization energy is associated with minigaps induced by lattice distortion in the $3d_{z^2}$ and $3d_{xz/yz}$ bands, which open a pseudogap in the $3d_{z^2}$ DOS and a small gap in the $3d_{xz/yz}$ dispersion.

In contrast to the cuprates, LaNiO$_2$ differs from the insulating cuprate parent compounds in that the former is metal, with localized Ni 3$d_{z^2}$, itinerant 3$d_{xz/yz}$, and La 5$d$ hybridized bands, indicating that a multiorbital model is necessary to properly describe the role of valence fluctuations between Ni$^{2+}$ ($d^8$) and Ni$^{1+}$ ($d^9$). The  behavior of the magnetic state is similar to the cuprates, except for a somewhat larger energy scale. That is, the magnetic gap in the Ni-$d_{x^2-y^2}$ band is approximately double that in the cuprates owing to the larger magnetic moments. Consequently, the energetic ordering scale of magnetic and stripe states in Fig.~\ref{fig:fig2}(b) is a factor of 5-7 larger than those in YBCO$_7$~\cite{Zhang2020}.

\section*{Discussion}
Despite significant differences between the cuprates and nickelates, there is a great deal of similarity between the results of Fig.~\ref{fig:fig2} and those of Figs. 2 and 3 for YBCO$_7$ in ref.~\cite{Zhang2020}. We suggest that this similarity is a signature of strong correlations being at play in correlated materials that are realized as a form of short-range order in which topological defects--here domain walls--play a significant role.  Topological defects are also known to play an important role in Martensitic phases~\cite{KarthaPRB1995} and in quenches~\cite{ThiOxford2016}.

While we have found 7 stripe phases with energies comparable to or lower than the $C$-AFM phase, there are actually an infinite number of such phases.  As $P_c$ gets larger, the stripe phase energy has to converge to the $C$-AFM energy. If we think of a phase as a pole of some susceptibility, we have demonstrated that nickelates and cuprates contain accumulation points of poles.  In complex analysis, an accumulation point of the poles of a holomorphic function forms a natural boundary for that function, beyond which it cannot be analytically continued. While such notions have not been addressed in band structure theory, we note that recent calculations have revived McMillan's notion that strongly correlated materials are associated with large bosonic entropy arising from many competing phases in the susceptibility~\cite{markiewicz2017entropic,McmillanPRB1977}.

Our findings support the experimental observation of fluctuating magnetic effects in nickelates~\cite{Hayward1999,Hayward2003,ZhaoPRL2021,Lu2021science,2021arXiv211202484R,Krieger2021,Tam2021,OrtizPRR2022,Leonov2020,Kitatani2020}. We find nearly degenerate orders of bond-centered stripes and $C$-AFM phases, along with flat bands in the low-energy region around the Fermi energy. Additional competing spin and charge orders are suggested by our many-body theory calculations, which could lead to intertwined orders in pristine LaNiO$_{2}$. Spin fluctuations associated with such nearly degenerate states could play a significant role in driving superconductivity, as suggested for other unconventional superconductors~\cite{Kievelson2003,Eduardo2015,Elbio2005}. A doping-dependent study is needed to fully elucidate this connection. 

The CDW predicted by our calculations agrees well with the experimental observations~\cite{2021arXiv211202484R,Krieger2021,Tam2021}. Furthermore, the stripe phases are stabilized by inducing Peierls distortion, leading to the minigaps in the $3d_{z^2}$ and $3d_{xz/yz}$ bands. Our findings strongly indicate that \textit{e-ph} coupling is important in  \textit{undoped} LaNiO$_2$, and the low-energy  physics of the nickelates is governed by multi-orbitals, consistent with the experimental findings in ref.~\cite{Tam2021}, and considerable \textit{e-ph} coupling effects are found in ref.~\cite{Gao2022magstrain}. Although the interplay between CDW and superconductivity needs further theoretical and experimental studies, our DFT and many-body calculations provide strong evidence of a striking similarity between the {\it undoped} nickelates and {\it doped} cuprates, both hosting strong correlation and \textit{e-ph} couplings effects.

\section*{Conclusion}

Using first-principles calculations and many-body theory based modeling, we show that the mechanism driving the formation of CDWs in the pristine infinite-layer nickelates is the presence of a strong Peierls instability involving several Ni $d$-orbitals, which is quite distinct from the case of the cuprates. In addition to the electronic mechanisms suggested by~\cite{2021arXiv211202484R}, our analysis identifies the key role of electron-phonon effects in the CDW phase. Importantly, the Ni 3$d_{z^2}$ orbital is found to lie close to the Fermi energy, indicating that it could play an important role in superconductivity and that a multi-orbital minimal model is needed to capture the rich physics of the nickelates.

\bibliography{Ref}

\begin{thebibliography}{10}
\expandafter\ifx\csname url\endcsname\relax
  \def\url#1{\texttt{#1}}\fi
\expandafter\ifx\csname urlprefix\endcsname\relax\def\urlprefix{URL }\fi
\providecommand{\bibinfo}[2]{#2}
\providecommand{\eprint}[2][]{\url{#2}}

\bibitem{Li2019a}
\bibinfo{author}{Li, D.} \emph{et~al.}
\newblock \bibinfo{title}{{Superconductivity in an infinite-layer nickelate}}.
\newblock \emph{\bibinfo{journal}{Nature}} \textbf{\bibinfo{volume}{572}},
  \bibinfo{pages}{624--627} (\bibinfo{year}{2019}).

\bibitem{Osada2020a}
\bibinfo{author}{Osada, M.} \emph{et~al.}
\newblock \bibinfo{title}{{A Superconducting Praseodymium Nickelate with
  Infinite Layer Structure}}.
\newblock \emph{\bibinfo{journal}{Nano Letters}} \textbf{\bibinfo{volume}{20}},
  \bibinfo{pages}{5735--5740} (\bibinfo{year}{2020}).

\bibitem{ZhangS2021}
\bibinfo{author}{Zeng, S.} \emph{et~al.}
\newblock \bibinfo{title}{{Superconductivity in infinite-layer nickelate
  La$_{1_x}$Ca$_{x}$NiO$_{2}$}}.
\newblock \emph{\bibinfo{journal}{Sci. Adv.}} \textbf{\bibinfo{volume}{8}},
  \bibinfo{pages}{eabl9927} (\bibinfo{year}{2022}).

\bibitem{2021arXiv211202484R}
\bibinfo{author}{Rossi, M.} \emph{et~al.}
\newblock \bibinfo{title}{{A broken translational symmetry state in an
  infinite-layer nickelate}}.
\newblock \emph{\bibinfo{journal}{Nat. Phys.}} \textbf{\bibinfo{volume}{18}},
  \bibinfo{pages}{869--873} (\bibinfo{year}{2022}).

\bibitem{OsadaAM2021}
\bibinfo{author}{Osada, M.} \emph{et~al.}
\newblock \bibinfo{title}{Nickelate superconductivity without rare-earth
  magnetism: (la,sr)nio2}.
\newblock \emph{\bibinfo{journal}{Adv. Mater.}} \textbf{\bibinfo{volume}{33}},
  \bibinfo{pages}{2104083} (\bibinfo{year}{2021}).

\bibitem{Sawatzky2019}
\bibinfo{author}{Sawatzky, G.~A.}
\newblock \bibinfo{title}{{Superconductivity seen in a non-magnetic nickel
  oxide}}.
\newblock \emph{\bibinfo{journal}{Nature}} \textbf{\bibinfo{volume}{572}},
  \bibinfo{pages}{592--593} (\bibinfo{year}{2019}).

\bibitem{Norman2020}
\bibinfo{author}{Norman, M.~R.}
\newblock \bibinfo{title}{{Entering the Nickel Age of Superconductivity}}.
\newblock \emph{\bibinfo{journal}{Physics}} \textbf{\bibinfo{volume}{13}},
  \bibinfo{pages}{85} (\bibinfo{year}{2020}).

\bibitem{Pickett2021}
\bibinfo{author}{Pickett, W.~E.}
\newblock \bibinfo{title}{{The dawn of the nickel age of superconductivity}}.
\newblock \emph{\bibinfo{journal}{Nat. Rev. Phys.}}
  \textbf{\bibinfo{volume}{3}}, \bibinfo{pages}{7--8} (\bibinfo{year}{2021}).

\bibitem{Mitchell2021}
\bibinfo{author}{Mitchell, J.~F.}
\newblock \bibinfo{title}{{A Nickelate Renaissance}}.
\newblock \emph{\bibinfo{journal}{Front. Phys.}} \textbf{\bibinfo{volume}{9}},
  \bibinfo{pages}{1--8} (\bibinfo{year}{2021}).

\bibitem{Hepting2020}
\bibinfo{author}{Hepting, M.} \emph{et~al.}
\newblock \bibinfo{title}{{Electronic structure of the parent compound of
  superconducting infinite-layer nickelates}}.
\newblock \emph{\bibinfo{journal}{Nat. Mater.}} \textbf{\bibinfo{volume}{19}},
  \bibinfo{pages}{381--385} (\bibinfo{year}{2020}).

\bibitem{Lee2004}
\bibinfo{author}{Lee, K.-W.} \& \bibinfo{author}{Pickett, W.~E.}
\newblock \bibinfo{title}{{Infinite-layer
  $\mathrm{La}\mathrm{Ni}{\mathrm{O}}_{2}$: ${\mathrm{Ni}}^{1+}$ is not
  ${\mathrm{Cu}}^{2+}$}}.
\newblock \emph{\bibinfo{journal}{Phys. Rev. B}} \textbf{\bibinfo{volume}{70}},
  \bibinfo{pages}{165109} (\bibinfo{year}{2004}).

\bibitem{Zhang2020d}
\bibinfo{author}{Li, D.} \emph{et~al.}
\newblock \bibinfo{title}{Superconducting dome in
  ${\mathrm{nd}}_{1\ensuremath{-}x}{\mathrm{sr}}_{x}{\mathrm{nio}}_{2}$
  infinite layer films}.
\newblock \emph{\bibinfo{journal}{Phys. Rev. Lett.}}
  \textbf{\bibinfo{volume}{125}}, \bibinfo{pages}{027001}
  (\bibinfo{year}{2020}).

\bibitem{Sakakibara2019}
\bibinfo{author}{Sakakibara, H.} \emph{et~al.}
\newblock \bibinfo{title}{Model construction and a possibility of cupratelike
  pairing in a new ${d}^{9}$ nickelate superconductor
  $(\mathrm{Nd},\mathrm{Sr}){\mathrm{nio}}_{2}$}.
\newblock \emph{\bibinfo{journal}{Phys. Rev. Lett.}}
  \textbf{\bibinfo{volume}{125}}, \bibinfo{pages}{077003}
  (\bibinfo{year}{2020}).

\bibitem{Botana2020}
\bibinfo{author}{Botana, A.~S.} \& \bibinfo{author}{Norman, M.~R.}
\newblock \bibinfo{title}{Similarities and differences between
  ${\mathrm{lanio}}_{2}$ and ${\mathrm{cacuo}}_{2}$ and implications for
  superconductivity}.
\newblock \emph{\bibinfo{journal}{Phys. Rev. X}} \textbf{\bibinfo{volume}{10}},
  \bibinfo{pages}{011024} (\bibinfo{year}{2020}).

\bibitem{Goodge2021}
\bibinfo{author}{Goodge, B.~H.} \emph{et~al.}
\newblock \bibinfo{title}{{Doping evolution of the Mott–Hubbard landscape in
  infinite-layer nickelates}}.
\newblock \emph{\bibinfo{journal}{Proc. Natl. Acad. Sci. U.S.A.}}
  \textbf{\bibinfo{volume}{118}}, \bibinfo{pages}{e2007683118}
  (\bibinfo{year}{2021}).

\bibitem{Jiang2020}
\bibinfo{author}{Jiang, M.}, \bibinfo{author}{Berciu, M.} \&
  \bibinfo{author}{Sawatzky, G.~A.}
\newblock \bibinfo{title}{{Critical Nature of the Ni Spin State in Doped
  ${\mathrm{NdNiO}}_{2}$}}.
\newblock \emph{\bibinfo{journal}{Phys. Rev. Lett.}}
  \textbf{\bibinfo{volume}{124}}, \bibinfo{pages}{207004}
  (\bibinfo{year}{2020}).

\bibitem{Zhang2021}
\bibinfo{author}{Zhang, R.} \emph{et~al.}
\newblock \bibinfo{title}{{Magnetic and f-electron effects in LaNiO$_{2}$ and
  NdNiO$_{2}$ nickelates with cuprate-like 3$d_{x^2-y^2}$ band}}.
\newblock \emph{\bibinfo{journal}{Commun. Phys.}} \textbf{\bibinfo{volume}{4}},
  \bibinfo{pages}{118} (\bibinfo{year}{2021}).

\bibitem{Lu2021science}
\bibinfo{author}{Lu, H.} \emph{et~al.}
\newblock \bibinfo{title}{Magnetic excitations in infinite-layer nickelates}.
\newblock \emph{\bibinfo{journal}{Science}} \textbf{\bibinfo{volume}{373}},
  \bibinfo{pages}{213--216} (\bibinfo{year}{2021}).

\bibitem{Fowlie2022}
\bibinfo{author}{Fowlie, J.} \emph{et~al.}
\newblock \bibinfo{title}{{Intrinsic magnetism in superconducting
  infinite-layer nickelates}}.
\newblock \emph{\bibinfo{journal}{Nat. Phys.}} \textbf{\bibinfo{volume}{18}},
  \bibinfo{pages}{1043--1047} (\bibinfo{year}{2022}).

\bibitem{Lin_2022}
\bibinfo{author}{Lin, H.} \emph{et~al.}
\newblock \bibinfo{title}{{Universal spin-glass behaviour in bulk LaNiO$_2$,
  PrNiO$_2$ and NdNiO$_2$}}.
\newblock \emph{\bibinfo{journal}{New J. Phys.}} \textbf{\bibinfo{volume}{24}},
  \bibinfo{pages}{013022} (\bibinfo{year}{2022}).

\bibitem{OrtizPRR2022}
\bibinfo{author}{Ortiz, R.~A.} \emph{et~al.}
\newblock \bibinfo{title}{Magnetic correlations in infinite-layer nickelates:
  An experimental and theoretical multimethod study}.
\newblock \emph{\bibinfo{journal}{Phys. Rev. Research}}
  \textbf{\bibinfo{volume}{4}}, \bibinfo{pages}{023093} (\bibinfo{year}{2022}).

\bibitem{Krieger2021}
\bibinfo{author}{Krieger, G.} \emph{et~al.}
\newblock \bibinfo{title}{Charge and spin order dichotomy in
  ${\mathrm{ndnio}}_{2}$ driven by the capping layer}.
\newblock \emph{\bibinfo{journal}{Phys. Rev. Lett.}}
  \textbf{\bibinfo{volume}{129}}, \bibinfo{pages}{027002}
  (\bibinfo{year}{2022}).

\bibitem{Tam2021}
\bibinfo{author}{Tam, C.~C.} \emph{et~al.}
\newblock \bibinfo{title}{{Charge density waves in infinite-layer NdNiO$_{2}$
  nickelates}}.
\newblock \emph{\bibinfo{journal}{Nat. Mater.}} \textbf{\bibinfo{volume}{21}},
  \bibinfo{pages}{1116--1120} (\bibinfo{year}{2022}).
\newblock \eprint{2112.04440}.

\bibitem{Gao2022magstrain}
\bibinfo{author}{{Gao}, Q.} \emph{et~al.}
\newblock \bibinfo{title}{{Magnetic Excitations in Strained Infinite-layer
  Nickelate PrNiO2}}.
\newblock \emph{\bibinfo{journal}{arXiv e-prints}}
  \bibinfo{pages}{arXiv:2208.05614} (\bibinfo{year}{2022}).
\newblock \eprint{2208.05614}.

\bibitem{Zhang2020}
\bibinfo{author}{Zhang, Y.} \emph{et~al.}
\newblock \bibinfo{title}{{Competing stripe and magnetic phases in the cuprates
  from first principles}}.
\newblock \emph{\bibinfo{journal}{Proc. Natl. Acad. Sci. U.S.A.}}
  \textbf{\bibinfo{volume}{117}}, \bibinfo{pages}{68--72}
  (\bibinfo{year}{2020}).

\bibitem{ChoiPRR2020}
\bibinfo{author}{Choi, M.-Y.}, \bibinfo{author}{Pickett, W.~E.} \&
  \bibinfo{author}{Lee, K.-W.}
\newblock \bibinfo{title}{Fluctuation-frustrated flat band instabilities in
  ${\mathrm{ndnio}}_{2}$}.
\newblock \emph{\bibinfo{journal}{Phys. Rev. Research}}
  \textbf{\bibinfo{volume}{2}}, \bibinfo{pages}{033445} (\bibinfo{year}{2020}).

\bibitem{ChrisLCO2018}
\bibinfo{author}{Lane, C.} \emph{et~al.}
\newblock \bibinfo{title}{Antiferromagnetic ground state of
  ${\mathrm{la}}_{2}{\mathrm{cuo}}_{4}$: A parameter-free ab initio
  description}.
\newblock \emph{\bibinfo{journal}{Phys. Rev. B}} \textbf{\bibinfo{volume}{98}},
  \bibinfo{pages}{125140} (\bibinfo{year}{2018}).

\bibitem{LechermannPRB2022}
\bibinfo{author}{Lechermann, F.}
\newblock \bibinfo{title}{Emergent flat-band physics in
  ${d}^{9\ensuremath{-}\ensuremath{\delta}}$ multilayer nickelates}.
\newblock \emph{\bibinfo{journal}{Phys. Rev. B}}
  \textbf{\bibinfo{volume}{105}}, \bibinfo{pages}{155109}
  (\bibinfo{year}{2022}).

\bibitem{Robert2021}
\bibinfo{author}{{Markiewicz}, R.~S.}, \bibinfo{author}{{Singh}, B.},
  \bibinfo{author}{{Lane}, C.} \& \bibinfo{author}{{Bansil}, A.}
\newblock \bibinfo{title}{{High-order Van Hove singularities in cuprates and
  related high-Tc superconductors}}.
\newblock \emph{\bibinfo{journal}{arXiv e-prints}}
  \bibinfo{pages}{arXiv:2105.04546} (\bibinfo{year}{2021}).
\newblock \eprint{2105.04546}.

\bibitem{MarkiewiczPRB2000}
\bibinfo{author}{Markiewicz, R.~S.}
\newblock \bibinfo{title}{Dispersion of ordered stripe phases in the cuprates}.
\newblock \emph{\bibinfo{journal}{Phys. Rev. B}} \textbf{\bibinfo{volume}{62}},
  \bibinfo{pages}{1252--1269} (\bibinfo{year}{2000}).

\bibitem{KarthaPRB1995}
\bibinfo{author}{Kartha, S.}, \bibinfo{author}{Krumhansl, J.~A.},
  \bibinfo{author}{Sethna, J.~P.} \& \bibinfo{author}{Wickham, L.~K.}
\newblock \bibinfo{title}{Disorder-driven pretransitional tweed pattern in
  martensitic transformations}.
\newblock \emph{\bibinfo{journal}{Phys. Rev. B}} \textbf{\bibinfo{volume}{52}},
  \bibinfo{pages}{803--822} (\bibinfo{year}{1995}).

\bibitem{ThiOxford2016}
\bibinfo{author}{Giamarchi, T.}, \bibinfo{author}{Millis, A.},
  \bibinfo{author}{Parcollet, O.}, \bibinfo{author}{Saleur, H.} \&
  \bibinfo{author}{Cugliandolo, L.}
\newblock \emph{\bibinfo{title}{Strongly Interacting Quantum Systems out of
  Equilibrium}} (\bibinfo{publisher}{Oxford University Press},
  \bibinfo{year}{2016}).

\bibitem{markiewicz2017entropic}
\bibinfo{author}{Markiewicz, R.}, \bibinfo{author}{Buda, I.},
  \bibinfo{author}{Mistark, P.}, \bibinfo{author}{Lane, C.} \&
  \bibinfo{author}{Bansil, A.}
\newblock \bibinfo{title}{Entropic origin of pseudogap physics and a
  mott-slater transition in cuprates}.
\newblock \emph{\bibinfo{journal}{Sci. Rep.}} \textbf{\bibinfo{volume}{7}},
  \bibinfo{pages}{1--21} (\bibinfo{year}{2017}).

\bibitem{McmillanPRB1977}
\bibinfo{author}{McMillan, W.~L.}
\newblock \bibinfo{title}{Microscopic model of charge-density waves in
  $2h\ensuremath{-}\mathrm{Ta}{\mathrm{se}}_{2}$}.
\newblock \emph{\bibinfo{journal}{Phys. Rev. B}} \textbf{\bibinfo{volume}{16}},
  \bibinfo{pages}{643--650} (\bibinfo{year}{1977}).

\bibitem{Hayward1999}
\bibinfo{author}{Hayward, M.~A.}, \bibinfo{author}{Green, M.~A.},
  \bibinfo{author}{Rosseinsky, M.~J.} \& \bibinfo{author}{Sloan, J.}
\newblock \bibinfo{title}{{Sodium Hydride as a Powerful Reducing Agent for
  Topotactic Oxide Deintercalation: Synthesis and Characterization of the
  Nickel(I) Oxide LaNiO$_{2}$}}.
\newblock \emph{\bibinfo{journal}{J. Am. Chem. Soc.}}
  \textbf{\bibinfo{volume}{121}}, \bibinfo{pages}{8843--8854}
  (\bibinfo{year}{1999}).

\bibitem{Hayward2003}
\bibinfo{author}{Hayward, M.} \& \bibinfo{author}{Rosseinsky, M.}
\newblock \bibinfo{title}{{Synthesis of the infinite layer Ni(I) phase
  NdNiO$_{2+x}$ by low temperature reduction of NdNiO$_{3}$ with sodium
  hydride}}.
\newblock \emph{\bibinfo{journal}{Solid State Sci.}}
  \textbf{\bibinfo{volume}{5}}, \bibinfo{pages}{839--850}
  (\bibinfo{year}{2003}).

\bibitem{ZhaoPRL2021}
\bibinfo{author}{Zhao, D.} \emph{et~al.}
\newblock \bibinfo{title}{Intrinsic spin susceptibility and pseudogaplike
  behavior in infinite-layer ${\mathrm{lanio}}_{2}$}.
\newblock \emph{\bibinfo{journal}{Phys. Rev. Lett.}}
  \textbf{\bibinfo{volume}{126}}, \bibinfo{pages}{197001}
  (\bibinfo{year}{2021}).

\bibitem{Leonov2020}
\bibinfo{author}{Leonov, I.}, \bibinfo{author}{Skornyakov, S.~L.} \&
  \bibinfo{author}{Savrasov, S.~Y.}
\newblock \bibinfo{title}{Lifshitz transition and frustration of magnetic
  moments in infinite-layer ${\mathrm{ndnio}}_{2}$ upon hole doping}.
\newblock \emph{\bibinfo{journal}{Phys. Rev. B}}
  \textbf{\bibinfo{volume}{101}}, \bibinfo{pages}{241108}
  (\bibinfo{year}{2020}).

\bibitem{Kitatani2020}
\bibinfo{author}{Kitatani, M.} \emph{et~al.}
\newblock \bibinfo{title}{{Nickelate superconductors—a renaissance of the
  one-band Hubbard model}}.
\newblock \emph{\bibinfo{journal}{npj Quantum Mater.}}
  \textbf{\bibinfo{volume}{5}}, \bibinfo{pages}{59} (\bibinfo{year}{2020}).

\bibitem{Kievelson2003}
\bibinfo{author}{Kivelson, S.~A.} \emph{et~al.}
\newblock \bibinfo{title}{How to detect fluctuating stripes in the
  high-temperature superconductors}.
\newblock \emph{\bibinfo{journal}{Rev. Mod. Phys.}}
  \textbf{\bibinfo{volume}{75}}, \bibinfo{pages}{1201--1241}
  (\bibinfo{year}{2003}).

\bibitem{Eduardo2015}
\bibinfo{author}{Fradkin, E.}, \bibinfo{author}{Kivelson, S.~A.} \&
  \bibinfo{author}{Tranquada, J.~M.}
\newblock \bibinfo{title}{Colloquium: Theory of intertwined orders in high
  temperature superconductors}.
\newblock \emph{\bibinfo{journal}{Rev. Mod. Phys.}}
  \textbf{\bibinfo{volume}{87}}, \bibinfo{pages}{457--482}
  (\bibinfo{year}{2015}).

\bibitem{Elbio2005}
\bibinfo{author}{Dagotto, E.}
\newblock \bibinfo{title}{{Complexity in Strongly Correlated Electronic
  Systems}}.
\newblock \emph{\bibinfo{journal}{Science}} \textbf{\bibinfo{volume}{309}},
  \bibinfo{pages}{257--262} (\bibinfo{year}{2005}).

\bibitem{Kresse1999}
\bibinfo{author}{Kresse, G.} \& \bibinfo{author}{Joubert, D.}
\newblock \bibinfo{title}{{From ultrasoft pseudopotentials to the projector
  augmented-wave method}}.
\newblock \emph{\bibinfo{journal}{Phys. Rev. B}} \textbf{\bibinfo{volume}{59}},
  \bibinfo{pages}{1758--1775} (\bibinfo{year}{1999}).

\bibitem{Kresse1993}
\bibinfo{author}{Kresse, G.} \& \bibinfo{author}{Hafner, J.}
\newblock \bibinfo{title}{{Ab initio molecular dynamics for open-shell
  transition metals}}.
\newblock \emph{\bibinfo{journal}{Phys. Rev. B}} \textbf{\bibinfo{volume}{48}},
  \bibinfo{pages}{13115--13118} (\bibinfo{year}{1993}).

\bibitem{Kresse1996}
\bibinfo{author}{Kresse, G.} \& \bibinfo{author}{Furthm{\"{u}}ller, J.}
\newblock \bibinfo{title}{{Efficient iterative schemes for ab initio
  total-energy calculations using a plane-wave basis set}}.
\newblock \emph{\bibinfo{journal}{Phys. Rev. B}} \textbf{\bibinfo{volume}{54}},
  \bibinfo{pages}{11169--11186} (\bibinfo{year}{1996}).

\bibitem{Sun2015}
\bibinfo{author}{Sun, J.}, \bibinfo{author}{Ruzsinszky, A.} \&
  \bibinfo{author}{Perdew, J.}
\newblock \bibinfo{title}{{Strongly Constrained and Appropriately Normed
  Semilocal Density Functional}}.
\newblock \emph{\bibinfo{journal}{Phys. Rev. Lett.}}
  \textbf{\bibinfo{volume}{115}}, \bibinfo{pages}{036402}
  (\bibinfo{year}{2015}).

\bibitem{ZhangPhysRevB2020}
\bibinfo{author}{Zhang, Y.} \emph{et~al.}
\newblock \bibinfo{title}{Symmetry-breaking polymorphous descriptions for
  correlated materials without interelectronic u}.
\newblock \emph{\bibinfo{journal}{Phys. Rev. B}}
  \textbf{\bibinfo{volume}{102}}, \bibinfo{pages}{045112}
  (\bibinfo{year}{2020}).

\bibitem{Furness2018}
\bibinfo{author}{Furness, J.~W.} \emph{et~al.}
\newblock \bibinfo{title}{{An accurate first-principles treatment of
  doping-dependent electronic structure of high-temperature cuprate
  superconductors}}.
\newblock \emph{\bibinfo{journal}{Commun. Phys.}} \textbf{\bibinfo{volume}{1}},
  \bibinfo{pages}{11} (\bibinfo{year}{2018}).

\bibitem{ZhangPhysRevB2022}
\bibinfo{author}{Zhang, R.} \emph{et~al.}
\newblock \bibinfo{title}{Critical role of magnetic moments in heavy-fermion
  materials: Revisiting ${\mathrm{smb}}_{6}$}.
\newblock \emph{\bibinfo{journal}{Phys. Rev. B}}
  \textbf{\bibinfo{volume}{105}}, \bibinfo{pages}{195134}
  (\bibinfo{year}{2022}).

\bibitem{Popescu2012}
\bibinfo{author}{Popescu, V.} \& \bibinfo{author}{Zunger, A.}
\newblock \bibinfo{title}{{Extracting $E$ versus $\Vec{k}$ effective band
  structure from supercell calculations on alloys and impurities}}.
\newblock \emph{\bibinfo{journal}{Phys. Rev. B}} \textbf{\bibinfo{volume}{85}},
  \bibinfo{pages}{085201} (\bibinfo{year}{2012}).

\end{thebibliography}

\section{Methods}
First principles calculations were performed by using the pseudopotential projector-augmented wave method~\cite{Kresse1999} as implemented in the Vienna {\it ab initio} simulation package (VASP)~\cite{Kresse1993,Kresse1996}. A high-energy cutoff of 520 eV was used to truncate the plane-wave basis set. Exchange-correlation effects were treated using the strongly-constrained-and-appropriately-normed (SCAN)~\cite{Sun2015} density functional. Due to the reduction of self-interaction error, SCAN improves over other convectional density functionals for a wide range of properties of correlated materials as reported in previous studies, e.g., for transition metal monoxides~\cite{ZhangPhysRevB2020}, cuprates~\cite{Zhang2020,Furness2018, ChrisLCO2018}, 
nickelates~\cite{Zhang2021}, and rare earth hexaborides~\cite{ZhangPhysRevB2022}. Crystal structures and ionic positions were fully optimized in all cases using a force convergence criterion of 0.01  eV/\AA{}  for each atom along with a total energy tolerance of 10$ ^{-5} $ eV.  A very dense $k$-point density (less than 0.02 $\AA^{-1}$) was used for relaxation of the structures. A denser $k$-point density of 0.015 $\AA^{-1}$ was used for total energy calculations. The unfolded band structures including orbital characters were extracted from the supercell pseudo-wavefunction calculations~\cite{Popescu2012}. The details of our further many-body calculations can be found in the SI.

\section*{Acknowledgements}     
R.Z. and J.S. acknowledge the support of the U.S. Office of Naval Research (ONR) Grant No. N00014-22-1-2673. The work at Tulane University was also supported by the start-up funding from Tulane University, the Cypress Computational Cluster at Tulane, the Extreme Science and Engineering Discovery Environment (XSEDE), and the National Energy Research Scientific Computing Center. The work at Northeastern University was supported by the US Department of Energy (DOE), Office of Science, Basic Energy Sciences Grant No. DE-SC0022216 (accurate modeling of complex magnetic states) and benefited from Northeastern University’s Advanced Scientific Computation Center and the Discovery Cluster and the National Energy Research Scientific Computing Center through DOE Grant No. DE-AC02-05CH11231. The work at Los Alamos National  Laboratory was carried out under the auspices of the US Department of Energy (DOE) National Nuclear Security Administration under Contract No. 89233218CNA000001. It was supported by the LANL LDRD Program, and in part by the Center for Integrated Nanotechnologies, a DOE BES user facility, in partnership with the LANL Institutional Computing Program for computational resources. B.B. acknowledges support from the COST Action CA16218. The work at TIFR Mumbai was supported by the Department of Atomic Energy of the Government of India under project number 12-R\&D-TFR-5.10-0100.

\section*{Author contributions}
R.Z., C.L., J.N, and B.S. performed computations. R.Z, C.L., B.B., R.S.M., A.B., and J.S. led the investigations and designed the computational approaches, and provided computational infrastructure. All authors analyzed the data and contributed to the writing of the manuscript.

\section{Data Availability Statement}

Full data for all figures is available from the authors by request.

\section*{Additional information}
The authors declare no competing financial interests. 

\end{document}